\documentclass[A4,12pt]{article} 
\usepackage{geometry,lipsum}
\usepackage[numbers]{natbib}

\usepackage{dcolumn}

\usepackage{xcolor}             
\usepackage{enumitem}			
\usepackage{hyperref}	        
\hypersetup{colorlinks=true,linkcolor=blue,citecolor=blue,breaklinks={true}} 
\usepackage[multiple]{footmisc}	
\usepackage{footnote}			

\usepackage{graphicx}
\graphicspath{{./FIG/}}
\DeclareGraphicsExtensions{.eps,.png,.jpg,.jpeg,.bmp,.gif,.pdf}
\usepackage{import}

\usepackage{tabularx,booktabs}	
\usepackage{multirow}			

\usepackage{algorithm}
\usepackage{algorithmic}

\usepackage{tikz}
\usepackage{pgf}
\usepackage{pgfplots}
\pgfplotsset{compat=newest}
\usepackage{csvsimple}
\usepackage{pgfplotstable}
\usepgfplotslibrary{units}
\usetikzlibrary{pgfplots.units}
\usetikzlibrary{calc}

\usepackage{amsmath}      
\usepackage{amssymb}	  
\usepackage{bm}		      
\usepackage{physics}	  
\usepackage{dsfont}		  
\newcommand{\R}{\mathbb{R}}		

\usepackage{amsthm}	  
\theoremstyle{definition}

\begin{document}
	
	\begin{center}
		\begin{LARGE}
			\bf
			Particle filtering of dynamical networks: Highlighting observability issues \vspace{1.0cm} 
		\end{LARGE}
		
		{\sc  Arthur N. Montanari},    \vspace{0.1cm}
		
		Graduate Program in Electrical Engineering of the Universidade Federal de Minas Gerais (UFMG), Avenida Ant\^onio Carlos 6627, 31270-901, Belo Horizonte, Minas Gerais, Brazil. \vspace{0.3cm}

		{\sc Luis A. Aguirre} \vspace{0.1cm}
		
		Departamento de Engenharia Eletr\^onica, UFMG.
		\vspace{0.5cm}
		
		\today
		
	\end{center}

	\section*{Abstract}
	In a network of high-dimensionality, it is not feasible to measure every single node. Thus, an important goal in the literature is to define the optimal choice of sensor nodes that provides a reliable state reconstruction of the network system state-space. This is an observability problem. In this paper, we propose a particle filtering (PF) framework as a way to assess observability properties of a dynamical network, where each node is composed by an individual dynamical system. The PF framework is applied on two benchmarks, networks of Kuramoto and R\"ossler oscillators, to investigate how the interplay between dynamics and topology impacts the network observability. Based on the numerical results, we conjecture that, when the network nodal dynamics are heterogeneous, better observability is conveyed for sets of sensor nodes that share some dynamical affinity to its neighbourhood. Moreover, we also investigate how the choice of an internal measured variable of a multidimensional sensor node affects the PF performance. The PF framework effectiveness as an observability measure is compared to a well-consolidated nonlinear observability metric for a small network case and some chaotic systems benchmarks.
	

	\date{\today}

\newpage
	
\begin{quotation}
	
\textbf{The high-dimensionality of network systems makes it unfeasible to measure every single node. Physically, it is not (yet) possible to measure every single neuron of the one hundred billion neurons present in a brain network. Likewise, it is not of economic interest to place a phasor measurement unit in every single electrical substation of a power system. Indeed, the problem of observability is recurrent in many network systems. Thus, it is only natural that in the past ten years, the literature has been flooded by innumerable methods that provide a minimum selection of sensor nodes that claim to be sufficient and/or necessary to render a network observable. However, a framework that allows one to compare the effectiveness and practicability of these methods is still missing. Moreover, the problem becomes intrinsically more complex when dealing with \textit{dynamical networks}\textemdash that is, networks whose nodes are individual dynamical systems. In this paper, we develop a framework based on Bayesian filtering as a way to assess the degree of observability conveyed by a given set of sensor nodes. This allows a further investigation of the interplay between observability, nodal dynamics and topology on a network system, as well as a means of comparison of different methods provided in the literature.}

\end{quotation}

\section{Introduction}
\label{sec:intro}


Introduced by Rudolf Kalman \cite{Kalman1959}, observability is a property that determines if the initial state of a dynamical system can be reconstructed solely based on input and output signals \cite{Chi-TsongChen1999}. The classic concept of observability, and\textemdash by duality\textemdash controllability, is based on a \textit{crisp} definition, i.e. the system is either observable (controllable) or not. 

Perhaps, a more practical question might be whether an observable system is \textit{almost unobservable}. To that end, this crisp concept was extended by metrics that quantify observability in a continuous manner, measuring \textit{how well} the system trajectory can be reconstructed from its observations \cite{Friedland1975,Aguirre1995,Summers2016}. This approach is useful to investigate which variables are more relevant to render a system observable for practical purposes, e.g. state reconstruction or system embedding\cite{Letellier2002,Letellier2005}. 

Since, in network systems, it is practically and numerically unfeasible to perform measurements on every single node, a selection of sensor nodes, in which both quantity and positioning must be optimal, is required. This is an observability problem. However, due to high dimensionality, classical observability metrics face serious numerical and scalability issues. To circumvent these challenges, Liu \textit{et al.} pioneering works in controllability \cite{Liu2011} and observability \cite{Liu2013c} took a graph-inspired approach, grounded on Lin's controllability (observability) definition \cite{Lin1974a,Chan1992}, to develop an algorithm that select the ``minimum'' set of driver (sensor) nodes under which a complex network is controllable (observable). Because of the intuitive representation of network systems by graph models and the high scalability of graph metrics, many following works \cite{Leitold2017,Posfai2013,Jia2013a,Nacher2013,Gao2014,Yan2015} embraced this graph-theoretical approach to study the controllability and observability in networks.


Faced with several different proposals of observability (controllability) methods, a recurrent goal in the literature revolves around the question whether a novel method really estimates a ``better'' set of sensor (driver) nodes than other methods, or if the estimated set at least renders a system observable (controllable) for practical purposes. However, a benchmark framework to compare and validate the efficacy of novel metrics in the literature is still absent, as pointed out in Ref. \cite{Carroll2018}.


For instance, it is arguable that works grounded on crisp definitions of observability (controllability), such as Lin's structural definition\cite{Lin1974a,Chan1992} and following works \cite{Liu2011,Liu2013c}, are not really apt to classify if a certain set of sensor nodes is really the best one or even if its feasible. Indeed, the pioneering methods of Liu \textit{et al.}\cite{Liu2011,Liu2013c} have been shown to underestimate the required set of sensor\cite{Haber2017,Letellier2018} and driver\cite{Gates2015,Leitold2017,Wang2017} nodes. A natural approach to circumvent this problem is to extend metrics that gradually quantify system observability (controllability), under a specific set of sensor (driver) nodes, to a network context \cite{Pasqualetti2013,Gu2015,Summers2016}.


On the other hand, most network systems are only investigated exclusively from its network topology, i.e. the corresponding adjacency matrix \cite{Leitold2017}, which is insufficient from an observability point of view \cite{Aguirre2018}. However, most networks used as benchmarks in \cite{Liu2011,Yuan2013a}, including food webs, regulatory networks, power grids, electronic circuits and neuronal networks, are in fact \textit{dynamical networks} (see Table 1 of Ref. \cite[Supplementary information]{Leitold2017}), i.e. a connected graph in which each node has an individual (multidimensional) dynamical system. Cowan \textit{et al.} \cite{Cowan2012} show that significantly different results could be derived by considering the presence of self-edges that dictate the independent dynamic behaviour of any given node. This includes a new level of depth to the observability problem of network systems, specially when nonlinear dynamics are considered\cite{Cowan2012,Pasqualetti2013,Gu2015,Summers2016,Su2017,Leitold2017,Haber2017,Aguirre2018,Letellier2018}. 

\begin{figure}
	\centering
	\includegraphics[width=0.5\columnwidth]{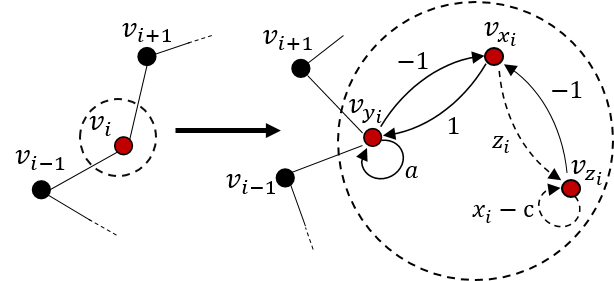} 
	\caption{Chain graph (left), and the corresponding graph representation when a given node $v_i$ is an individual R\"ossler system coupled by the $y$ variable \cite{Letellier2018,Aguirre2018}. Note that the R\"ossler system could be coupled by the $x$ or $z$ variable instead yielding very different results although the chain graph (topology in the left) would remain the same. Linear and nonlinear interactions are traced in solid and dashed lines, respectively.}
	\label{fig:rosslergraph}       
\end{figure}

As pointed out in Ref. \cite{Aguirre2018}, a dynamical network can be decomposed in three levels: i)~the nodal dynamics, described by an individual dynamical system in each node; ii)~the network topology, described by a graph; and iii)~the \textit{full network}, which is a combination of the aforementioned levels. Figure~\ref{fig:rosslergraph} illustrates the corresponding graph of a given network topology when nodal dynamics are considered. Despite recent investigations using small motifs \cite{Notarstefano2013,Whalen2015,Su2017}, the interplay between observability, nodal dynamics, especially with nonlinearities, and network topology is still not clear \cite{Haber2017,Aguirre2018}.

\subsection{Paper contributions}
\label{sec:introcontrib}

In this paper, we present a Bayesian filter formulation, based on the particle filter\footnote{Also known as sequential importance resampling (SIR) filter.} (PF), for dynamical network applications as a way to assess its observability properties. This is motivated by the fact that if the measured signals do not provide relevant information to the filter, i.e. they convey poor observability, then the update stage of the filter is impaired and, consequently, the estimates show poor performance. Thus, we argue that the provided PF formulation can be used as benchmark for observability studies, that is, as a means to validate and compare, on a practical level, different observability metrics and sensor selection methods proposed in the literature. Likewise, a moving horizon estimation technique \cite{Haber2017} as well as the fitting error from training a Gauss-Newton algorithm \cite{Guan2018} or a reservoir computer \cite{Carroll2018} have also been proposed in the literature as alternatives to assess the ``reconstruction quality'' of a dynamical system trajectory.

Having established a numerical framework based on the PF, this paper investigates a number of issues related to  the interplay between the network topology and nodal dynamics from an observability point of view. To this end two benchmark systems were used: a network of Kuramoto oscillators and a network of R\"ossler systems. In the former, every node is composed of a 1-dimensional model, the focus is exclusively on the impact of the {\it choice of sensor nodes}. In the latter, each node has a 3-dimensional model and this allows the study of the influence of sensor node selection and {\it the choice of dynamical variables at each sensor node}\, in terms of observability.  

The paper is organized as follows. Section \ref{sec:back} provides a background on observability metrics and PF. Section \ref{sec:setup} describes the PF methodology and numerical setup applied in each network benchmark, while Section \ref{sec:results} presents the numerical results. Section \ref{sec:conc} concludes the work.

\section{Background}
\label{sec:back}

\subsection{Particle filtering}
\label{sec:pf}

PF is a suboptimal state estimator based on sequential Monte Carlo methods to solve statistical problems. This method uses an independent number of stochastic samples, so-called \textit{particles}, extracted directly from the state-space, that are recursively located, propagated and weighted accordingly to Bayes' theorem. The method describes the states probability density function (PDF) by sampling approximation, not relying on any linearisation method or function approximation. Thus, it is applicable to any nonlinear and non-Gaussian problem, such as systems with several nonlinearities (e.g. chaotic systems) or multi-modal PDFs. The drawback is the computational burden, although the ever-increasing computational power already allows PF to be used in \textit{online} applications \cite{Doucet2011}, specially multi-target tracking \cite{Wang2017b}. 

This section provides background on the PF framework. The reader is referred to Refs. \cite{Doucet2000,Chen2003,Doucet2011,Candy2016a} for details. Let us consider the following discrete-time autonomous nonlinear dynamical system:
\begin{equation}
	\begin{aligned}
		\bm x_k &= f(\bm x_{k-1}) + \bm w_k \quad\quad && \text{state equation} \\
		\bm y_k &= h(\bm x_{k}) + \bm  v_k \quad\quad && \text{observation equation}
	\end{aligned}
	\label{eq:dynamicalsystem}
\end{equation}

\noindent
where $k=1,2,\dots$ are the time indices, $\bm x_k\in\R^n$ is the $n$-dimensional state at time instant $k$, $\bm y_k\in\R^q$ is the $q$-dimensional observation at time $k$, $f : \R^n\mapsto\R^n$ and $h : \R^n\mapsto\R^q$ are nonlinear functions, and $\bm  w_k\in\R^n$ and $\bm  v_k\in\R^q$ are independent additive process and observation noises, respectively with known distributions.

Given the system model \eqref{eq:dynamicalsystem} and a sequence of observations $\bm y_{1:k}=\left\lbrace \bm y_1, \bm y_2,\ldots,\bm y_k\right\rbrace$, the goal is to estimate the \textit{posterior} distribution $p(\bm x_k|\bm x_{1:k-1},\bm y_{1:k})$ sequentially, that is, from a \textit{prior} distribution $p(\bm x_{k-1}|\bm x_{1:k-2},\bm y_{1:k-1})$. It is assumed that model (\ref{eq:dynamicalsystem}) is a Markovian process, i.e. future states depend only upon the present state, yielding $p(\cdot | \bm x_{1:k-1}) = p(\cdot | \bm x_{k-1})$ and $p(\cdot | \bm y_{1:k}) = p(\cdot | \bm y_{k})$. Recall that the PF approximates a \textit{posterior} distribution $p(\bm x_k|\bm x_{k-1}, \bm y_{k})$ with random samples, which are propagated from time instant $k-1$ to $k$ by a mathematical model. The particles are assigned to weights based on the \textit{likelihood} distribution $p(\bm y_k|\bm x_k^{(p)})$ of the observations at time instant $k$, where $\bm x_k^{(p)}$ is a realization of the $p$-th particle. Finally, the filter output is the weighted average of all particle realizations. Algorithm~\ref{alg:pf} summarizes these proceedings.

\begin{figure}
\begin{algorithm}[H]
\begin{small}
\caption{Particle filter\label{alg:pf}}

For time steps $k=1,2,\dots$

\begin{enumerate}[noitemsep]

\item Given a \textit{proposal} distribution $q$, draw $N_p$ particles (samples) 
\begin{equation}
\bm x_k^{(p)}\sim q(\bm x_k^{(p)}|\bm x_{k-1}^{(p)},\bm y_k), \quad \forall p=1,2,\ldots,N_p ,
\end{equation} 

\noindent
where $\bm x_k^{(p)}$ is a realization of the $p$-th particle. 


\item Given a \textit{likelihood} distribution $p(\bm y_k|\bm x_k^{(p)})$ and a \textit{transition} distribution $p(\bm x_k^{(p)}|\bm x_{k-1}^{(p)})$, calculate the importance weights $W_k^{(p)}$ of the $p$-th particle at time $k$:
\begin{equation}
W_k^{(p)} = W_{k-1}^{(p)}\cdot \frac{p(\bm y_k|\bm x_k^{(p)}) \cdot p(\bm x_k^{(p)}|\bm x_{k-1}^{(p)})}{q(\bm x_k^{(p)}|\bm x_{k-1}^{(p)},\bm y_k)}, \quad \forall p.
\end{equation} 

\item Normalize the importance weights
\begin{equation}
\overline{W}_k^{(p)} = \frac{W_k^{(p)}}{\sum_{p=1}^{N_p} W_k^{(p)}}, \quad \forall p.
\end{equation}

\item Calculate the (estimated) effective sample size
\begin{equation}
\hat{N}_{\rm eff} = \frac{1}{\sum_{i=1}^{N_p} (\overline{W}_k^{(p)})^2}.
\end{equation}

\item If $\hat{N}_{\rm eff}<N_{\rm t}$, where $N_{\rm t}$ is a predefined threshold, resample $N_p$ particles and reset weights to $W_k^{(p)} = 1/N_p$.

\item Compute the filter output
\begin{equation}
	\hat{\bm x}_k = \frac{1}{N_p}\sum_{p=1}^{N_p} \bm x_k^{(p)}\overline{W}_k^{(p)},
\end{equation}

and increment time step $k$.

\end{enumerate}
\end{small}
\end{algorithm}
\end{figure}

Some remarks are in order: (i)~for resampling methods details, e.g. multinomial, residual and systematic, see Ref. \cite{Candy2016a}; (ii)~a common threshold\cite{Chen2003} is $N_{\rm t} = N_p/2$ or $N_p/3$; and (iii)~weights are usually calculated in logarithmic scale to improve numerical precision. Noise process must be carefully designed: a smaller noise process causes sample impoverishment in very few iterations, while higher noise process can provide estimates with larger deviations \cite{Chen2003}.

\subsection{Observability}
\label{sec:obs}

This section provides background on observability metrics for nonlinear systems. The reader is referred to Refs. \cite{Hermann1977,Letellier2005,Aguirre2005} for details. Let us consider the continuous-time autonomous nonlinear dynamical system
\begin{equation}
	\begin{aligned}
		\dot{\bm x} &= f(\bm x) \quad\quad && \text{state equation} \\
		\bm y &= h(\bm x)       \quad\quad && \text{observation equation}
	\end{aligned}
\label{eq:dynamicalsystemcont}
\end{equation}

\noindent
where $\bm x\in\R^n$, $\bm y\in\R^q$, $f:\R^n\mapsto\R^n$ and $h:\R^n\mapsto\R^q$.

The observability matrix of \eqref{eq:dynamicalsystemcont} can be formulated as
\begin{equation}
	\mathcal O_h(\bm x) = 
	\begin{bmatrix}
		\pdv{\mathcal L_{f}^{0}h(\bm x)}{\bm x} \\ \pdv{\mathcal L_{f}^{1}h(\bm x)}{\bm x} \\ \vdots \\ \pdv{\mathcal L_{f}^{n-1}h(\bm x)}{\bm x}
	\end{bmatrix} ,
\label{eq:nonlinearobservabilitymatrix}
\end{equation}

\noindent
where $\mathcal L^j_{f}h(\bm x)$ is the $j$th-order Lie derivative of $h$ along the vector field $f$. System \eqref{eq:dynamicalsystemcont} is said to be observable if \eqref{eq:nonlinearobservabilitymatrix} is full rank 
$\forall \bm x$, i.e. ${\rm rank}(\mathcal O_h(\bm x))=n,~\forall \bm x$. Otherwise, it is unobservable.

In order to quantify the system observability in a continuous manner, it is helpful to assess how far $\mathcal O_h(\bm x)$ is from being rank-deficient. Following the ideas in Refs. \cite{Friedland1975,Aguirre1995}, adapted to the nonlinear case\cite{Letellier1998,Letellier2002}, this can be achieved by computing a coefficient $\delta_o(\bm x)$ that quantifies the numerical ill-conditioning of $\mathcal O_h(\bm x)$ at a given state $\bm x$:
\begin{equation}
	\delta_o(\bm x) = \left|\frac{\lambda_{\rm min}\left(\mathcal O_h(\bm x)^\mathsf{T}\mathcal O_h(\bm x)\right)}{\lambda_{\rm max}\left(\mathcal O_h(\bm x)^\mathsf{T}\mathcal O_h(\bm x)\right)} \right| ,
\label{eq:observrank}
\end{equation}

\noindent
where $0\leq\delta_o(\bm x)\leq 1$ and $\lambda_{\rm max}(\cdot)$ indicates the maximum eigenvalue (likewise for $\lambda_{\rm min}$). In principle, the higher $\delta_o (\bm x)$, the more observable \eqref{eq:dynamicalsystemcont} is. If $\delta_o(\bm x)=0$, the system is unobservable. Averaging $\delta_o$ along a trajectory over interval $t\in\left[t_0,t_1\right]$ yields a ``global'' observability coefficient
\begin{equation}
\delta_o=\frac{1}{t_1-t_0}\int_{t_0}^{t_1}\delta_o(\bm x(\tau))\rm d\tau.
\label{eq:observrankglobal}
\end{equation}

\noindent
We note that the global coefficient $\delta_o$ is not absolute, but relative to the dynamical system. In other words, for a given system, two or more measuring functions can be investigated on the basis of their respective coefficient $\delta_o$. However, such observability coefficients of different systems are not necessarily comparable.  Check Ref. \cite{Aguirre2008} for details.

\section{Particle Filtering of Dynamical Networks}
\label{sec:setup}

A graph is defined as $\mathcal{G} = \{\mathcal{V},\mathcal{E},A\}$, where $\mathcal{V}=\{v_1,v_2,\dots,v_N\}$ and $\mathcal{E} \subseteq \mathcal{V} = \mathcal{V} \times \mathcal{V} = \{e_1,e_2,\dots,e_m\}$ are finite sets of $N$ nodes and $m$ edges, respectively. The adjacency matrix $A = [a_{ij}]$ is a mapping which associates elements (edges) of $\mathcal{E}$ to a pair of elements (nodes) of $\mathcal{V}$. If $(v_i,v_j)$ are unidirectionally linked, then $a_{ij}=a_{ji}=1$ (unweighted). Otherwise, $a_{ij} = 0$. The cardinality of a set is referred to as $\left|\mathcal V\right| = m$.

In the following experiments, there is a dynamical system at each node. A benchmark network is used to generate the data that will feed a particle filter. A hat, e.g. $\hat{\bm x}$, is used to indicate estimated values.

The computational framework is based on the PF for convenience. Despite the fact that PFs suffer with the inconvenient \textit{curse of dimensionality} \cite{Snyder2008,VanLeeuwen2015}, we deal only with networks with at most 45 states. Furthermore, as our second benchmark involves chaotic behaviour, the PF is seen to be a better alternative due to several nonlinearities and non-Gaussian uncertainties in the system \cite{Ching2006,Lingala2012}. We argue that Kalman filter-based estimators are not recommended in such cases since its covariance matrix is not computationally feasible in high-dimensional systems, especially those with high linear dependence among its states, such as network systems.

\subsection{Network of Kuramoto oscillators}
\label{sec:pfkuramoto}

Consider a network of Kuramoto oscillators where each oscillator $v_i\in\mathcal V$ is represented by a phase angle $x_i\in\R$. The dynamical benchmark network can be represented by the following continuous state-space model \cite{Wang2002,Chen2013}:
\begin{equation}
	\begin{aligned}
		\dot{x}_i = \omega_i + \rho\sum_{j = 1}^{N}a_{ij}\sin(x_j - x_i), \quad i=1,\ldots,N,
	\end{aligned}
\label{eq:kuramotonetwork}
\end{equation}

\noindent
where $N$ is the network size, $\rho$ is the coupling strength, and $\omega_i>0$ is the natural frequency of the $i$th oscillator. Hence the state vector of the full network is $\bm x = \begin{bmatrix} x_1 & x_2 & \dots & x_N \end{bmatrix}^\mathsf{T} \in\R^N$.
Note that the coupling among the oscillators is additive, diffusive and proportional to $\rho$. 

In the simulations $\rho=0.1$, $\omega_i\sim\mathcal{N}\left(1,0.03\right)$ and initial conditions $x_i(t_0)\sim\mathcal U(-\pi,+\pi)$, $\forall i$, where $\mathcal{N}\left(\mu,\sigma^2\right)$ denotes a Gaussian distribution with average $\mu$ and variance $\sigma^2$, and $\mathcal U\left(a,b\right)$ denotes a uniform distribution within limits $[a,\,b]$. Numerical integration is performed using a fourth-order Runge Kutta algorithm for  simulation time of $500$\,s, with time step $0.1$\,s.

The output vector is obtained as follows:
\begin{equation}
	\bm y_k = C \bm x_k + \bm v_k	,
	\label{eq:kuramotomeasure}
\end{equation}

\noindent
where $C\in\R^{q\times n}$ is the output matrix, $q$ is the number of sensor nodes, $\bm v_k\sim\mathcal N(0, \Sigma_v)$, $\Sigma_v = \sigma_v^2\cdot I_q$ is the covariance matrix of $\bm v_k$, $\sigma_v=0.1$ and $I_q$ is an identity matrix of dimension $q$. It is assumed that measures are taken independently. The set of sensor nodes is $\mathcal{S} = \{s_1,s_2,\ldots,s_q\} \subseteq \mathcal{V}$. A node  $v_j$ will be listed as a sensor node $s_i$ if $c_{ij}=1$, where $c_{ij}$ is an element of $C$. If measures are taken on all nodes, then $\mathcal{S} \equiv \mathcal{V}$.

The goal is to investigate how the selection of sensor nodes affects the PF performance. Other important aspects such as robustness to model parameter variation, noise level and dimensionality have been investigated elsewhere \cite{Chen2003,Snyder2008}. Thus, in the simulations the true values of $\omega_i$ will be used. 

The bootstrap PF is implemented as follows. Particles are propagated through a discrete numerical approximation $f(\bm x_{k-1}^{(p)}, \omega)$ of the continuous model (\ref{eq:kuramotonetwork}), obtained analytically via the backward Euler method with time step $0.1$ s.
We introduce an additive process noise $\bm w_k\sim\mathcal N\left(0, \Sigma_w\right)$\textemdash where $\Sigma_w=\sigma_w^2\cdot I_{n}$ is the covariance matrix of $\bm w_k$ and $\sigma_w=0.1$\textemdash so that the filter particles have a better coverage of the state-space, as well as avoiding sample impoverishment in few steps. We set $\hat{\bm x}(t_0) = \bm x(t_0)$ for fast convergence and $N_p=500$.

In this paper, we implement exclusively the \textit{bootstrap} PF variant, in which the proposal distribution is assumed to be equal to the transition distribution, yielding $q(\bm x_k^{(p)}|\bm x_{k-1}^{(p)},\bm y_k) = p(\bm x_k^{(p)}|\bm x_{k-1}^{(p)})$. Based on Algorithm \ref{alg:pf}, we summarize the implemented PF framework for the Kuramoto network in Algorithm \ref{alg:pfkuramoto}.

\begin{algorithm}[H]
\begin{small}
\caption{Bootstrap particle filter for the dynamical networks of Kuramoto oscillators and R\"ossler systems
	\label{alg:pfkuramoto}}

For time steps $k=1,2,\dots$

\begin{enumerate}[noitemsep]

\item Draw $N_p$ samples from 
$p(\bm x_{k}^{(p)}|\bm x_{k-1}^{(p)})$:
\begin{equation}
\bm x_{k}^{(p)} \sim \mathcal N \left(f(\bm x_{k-1}^{(p)}),\Sigma_w\right), \quad \, p=1,\ldots,N_p.
\end{equation}

\item Given a likelihood distribution $p(\bm y_k|\bm x_k^{(p)})\sim\mathcal N\left(\bm y_{k},\Sigma_v\right)$, calculate the importance weights $W_n^{(p)}$ according to
\end{enumerate}
\begin{equation}
\begin{aligned}
W_k^{(p)} &= W_{k-1}^{(p)}\cdot \frac{ \exp{-\frac{1}{2} (C\bm x_k^{(p)} - \bm y_k^{(p)})^\mathsf{T}\Sigma_v^{-1}(C\bm x_k^{(p)} - \bm y_k^{(p)})} }{(2\pi)^{\frac{q}{2}}\rm{det} (\Sigma_v)^{\frac{1}{2}}} \\
 &= W_{k-1}^{(p)}\cdot \frac{1}{(2\pi)^{q/2}\sigma_v^q}\cdot\exp{-\frac{1}{2}\norm{ \frac{C\bm x_k^{(p)} - \bm y_k}{\sigma_v}  }^2}, \,\forall p.
\end{aligned}
\end{equation}


\begin{enumerate}
\item[3.] Follow steps (3)--(6) from Algorithm \ref{alg:pf}, using a systematic resampling for $N_{\rm t} = 0.5$.
\end{enumerate}

\end{small}
\end{algorithm}

\subsection{Network of R\"ossler systems}
\label{sec:pfrossler}

Consider a network of R\"ossler oscillators, in a chaotic regime, coupled by the $y$ variable \cite{Pecora1990a,Boccaletti2002}. Hence, at each node $v_i\in\mathcal V$ there is a system with three state variables $\bm{x}_i = \begin{bmatrix} x_i & y_i & z_i \end{bmatrix}^\mathsf{T}$. The dynamical network can be represented by the following state-space model:
\begin{equation}
	\begin{aligned}
	\begin{cases}
		\dot{x}_i &= -y_i - z_i \\
		\dot{y}_i &= x_i + a_iy_i + \rho\sum_{j = 1}^{N}a_{ij}(y_j - y_i) \\
		\dot{z}_i &= b_i + z_i(x_i - c),
	\end{cases}
		, \quad i = 1,\dots,N
	\end{aligned}
\label{eq:rosslernetwork}
\end{equation}

\noindent
where  $N$ is the number of nodes, $n=3N$ is the number of state variables of the network and $(a_i,b_i,c_i)$ are the parameters of the $i$-th R\"ossler system. Numerical integration is performed using fourth-order Runge Kutta algorithm for a total simulation time of $500$ s, with time step $0.01$ s. Simulation parameters were set to $\rho=0.1$, initial conditions $\bm{x}_i(t_0)\sim\mathcal N(0, I_3)$, $(b_i,c_i) = (2,4)$, and $N$ equidistant values within interval $a_i\in[0.388,0.408], \forall i$. This corresponds to spiral dynamics for each node in the case of no coupling.

Measures are taken linearly on the network of oscillators as follows:
\begin{equation}
	\bm {y}_k = C \cdot \begin{bmatrix} \bm x_{1,k}^\mathsf{T} & \bm x_{2,k}^\mathsf{T} & \ldots & \bm x_{N,k}^\mathsf{T}  \end{bmatrix}^\mathsf{T} + \bm v_k	
	\label{eq:rosslermeasure}
\end{equation}

\noindent
where $C\in\R^{q\times n}$, $\bm v_k\sim\mathcal N(0, \Sigma_v)$ and $\sigma_v=0.1$. Note that the observation matrix $C$ defines the network sensor nodes as well as which variables are measured at each oscillator. Hence, given a sensor node, all variables could be measured $s_i = (x_i,y_i,z_i)$, or just one (e.g. $s_i = y_i$) or a combination of two variables.

Algorithm \ref{alg:pfkuramoto} summarizes the implemented PF for the R\"ossler network. The true values of $(a_i,b_i,c_i)$ are used. We set $N_p=500$, $\bm w_k\sim\mathcal N(0,\Sigma_w)$ and $\sigma_w=0.01$ for the reasons mentioned in Section \ref{sec:pfkuramoto}. Function $f(\bm x_{k-1}^{(p)}, a, b, c)$ is a discrete numerical approximation of the continuous model \eqref{eq:rosslernetwork} via the backward Euler method, with time step $0.01$ s.

\section{Numerical Results}
\label{sec:results}

%
%

The dynamical networks and PFs detailed in Section \ref{sec:setup} are considered in this section. Measures are taken independently at each  node and the network graph of interest is a \textit{chain graph} of $N$ nodes.

\subsection{Particle filtering framework as a validation benchmark}
\label{sec:pfbenchmark}

The quality of a Bayesian filter estimate depends on the observability conveyed by the measured variables. In other words, the higher the observability rank of a network system observed using a particular set of sensor nodes, the better the PF performance, that is, the smaller is the difference between its estimates and the true value. In this section, we analyze the PF framework effectiveness as an alternative mean to assess observability properties of a network system by comparing its performance with the results yielded by the well-consolidated observability rank in \eqref{eq:observrankglobal}. Since $\delta_o$ becomes unfeasible for high-order systems, we investigate the PF framework on a Kuramoto dynamical network of $N=4$ nodes. 

Moreover, it is important to notice that, unlike the observability metrics presented in Section \ref{sec:obs}, the PF is not a deterministic tool since its results heavily depend on the particle realizations. A statistical analysis is required. Thus, in the following numerical simulations, the applied performance index $\eta_i$ is the median of the normalized root-mean-square error\footnote{Simulations were carried out for different performance metrics, including the mean absolute error, mean absolute percentage error and mean relative error, yielded similar results.} (NRMSE) per node $v_i$ over 100 Monte Carlo simulations:
\begin{equation}
\eta_i = \text{median} \left\lbrace \frac{1}{\Delta\bm x_{i}} \sqrt{ \frac{1}{N}\sum_{k}\left(\bm x_{i,k} - \hat{\bm x}_{i,k|q}  \right)^2 } \right\rbrace ,
\end{equation} 

\noindent
where $ \hat{\bm x}_{i,k|q}$ refers to estimate $\hat{\bm x}_{i,k}$ of the $q$th Monte Carlo simulation (with different process noise realizations), and $\Delta\bm x_{i} = \max{\bm x_{i}} - \min{\bm x_{i}}$ is the measured data range. Note that median is preferred to the mean since the PDF of $\hat{\bm x}_{i,k}$ is non-Gaussian. The PF overall performance for a dynamical network is taken as $\eta = \frac{1}{n} \sum_{i=1}^{n}\eta_i$.

\begin{figure}
	\centering
	\includegraphics[width=0.5\columnwidth]{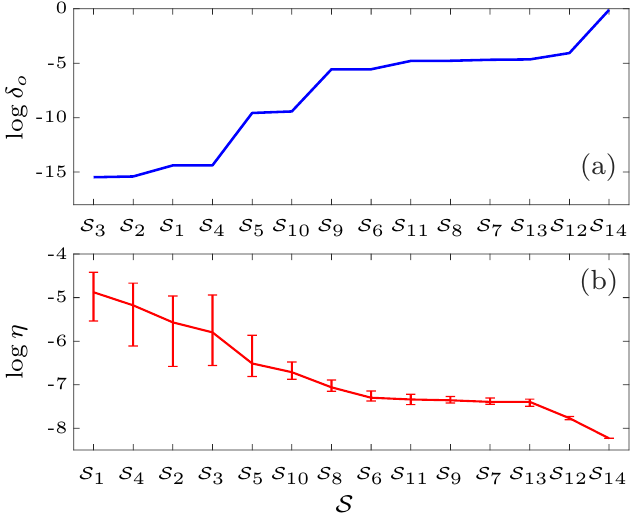}
	\caption{For every possible set of sensor nodes in a four-node network of Kuramoto oscillators, (a) the observability rank $\delta_o$ and (b) the performance index $\eta$, both in logarithmic scale. Error bars show the interquartile range. Sets $\mathcal S$ are sorted in crescent and decrescent order for $\delta_o$ and $\eta$, respectively. Simulations were performed with $\omega = \left[ 1.0155 \enspace 0.9648 \enspace 1.022 \enspace 1.0476 \right]^\mathsf{T}$. The sensor sets cardinalities are $|S_{1-4}|=1, |S_{5-10}|=2, |S_{11-13}|=3,$ and $|S_{14}|=4$.}
	\label{fig:obsmetrics}       
\end{figure}

Fig. \ref{fig:obsmetrics} compares the computed observability coefficients $\delta_o$ and the PF framework performance index $\eta$ for every possible combination of sensor nodes in this network. Firstly, it is interesting to note that \textit{all} choices of sensor nodes are observable (i.e. $\delta_o>0$ for all sets $\mathcal S$). This is in line with Liu \textit{et al.} structural approach \cite{Lin1974a,Liu2013c} since the whole graph is a root-strongly connected component (SCC). However, from a practical point-of-view, as pointed out by $\delta_o$ and $\eta$ metrics, some choices of sensor nodes (specially when $\left|S\right|=1$) are \textit{almost unobservable}. Thus, although Lin's structural definition of controllability (observability)\cite{Lin1974a,Chan1992} holds for \textit{almost every choice} of the parameter space (since it does not depend on the nonzero specific entries of the adjacency matrix), it might fail\textemdash for practical purposes\textemdash under specific entries\cite{Haber2017,Whalen2015}.

It is evident that the PF performance $\eta$ is not \textit{exactly} an equivalent metric to $\delta_o$. Specially for $|\mathcal S|=1$, where the correlation between $\eta$ and $\delta_o$ is the lowest. A possible justification is that, since systems of poor observability compromises the PF update stage efficacy, the PF realizations exhibits large interquartile ranges in its estimates which, in turn, also compromises the statistical relevance of $\eta$ for $|\mathcal S|$. In other words, if the error bars are taken into account the perfornace of the PF for all options with $|\mathcal S|=1$, it is seen that performance is almost equally poor from a statistical point of view.
However, when $|\mathcal S|\geq2$, we note that the standard deviation of the PF estimates $\eta$ reduces considerably and results become statistically relevant. Indeed, for $|\mathcal S|\geq2$, there is a high correlation between $\eta$ and $\delta_o$, except in two occasions ($\mathcal S_8$ and $\mathcal S_9$ are switched in Figs. \ref{fig:obsmetrics}a,b). We argue that although the PF is not effective to quantify or rank the observability of dynamical systems of very poor observability indexes, it is quite robust when the observability index is a little bit better. 

Appendix \ref{sec:chaoticsystems} further investigates the PF efficacy as an observability quantification in chaotic systems benchmarks explored in Ref. \cite{Aguirre2011a}.


\subsection{Network of Kuramoto oscillators}
\label{sec:kuramotoresults}

\begin{figure}
	\centering
	\includegraphics[width=0.45\columnwidth]{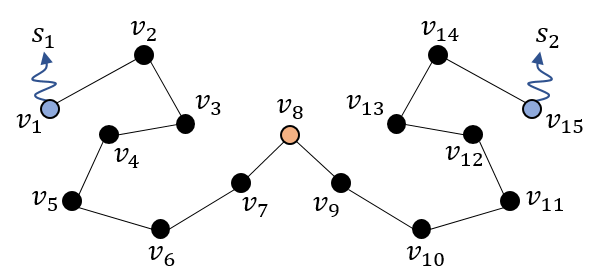}
	\caption{Chain graph of Kuramoto oscillators, with measures taken on network ends. Sensor nodes are marked in blue and the node with highest path length from all sensor nodes is marked in orange.}
	\label{fig:netchain}       
\end{figure}

Consider a Kuramoto dynamical network of $N=15$ nodes where measures are taken independently on the chain extremities ($\mathcal{S}_1=\{v_1,v_{15}\}$, see Fig. \ref{fig:netchain}). Thus:
\begin{equation}
	C = \begin{bmatrix} 
		1 & 0 & \ldots & 0 & 0 \\
		0 & 0 & \ldots & 0 & 1
		\end{bmatrix}.
\end{equation}

\noindent
where $C\in\R^{2\times 15}$. Following the results in Section \ref{sec:pfbenchmark}, it is implied that a node (state) with the worst PF estimate (highest $\eta_i$) is the least observable node from a particular set of sensor nodes. Hence, the optimal selection of sensor nodes is actually the one that provides minimum overall estimation error $\eta$ when compared to all other options with the same number of sensor nodes.

\begin{figure*}[t]
	\centering
	\includegraphics[width=\textwidth]{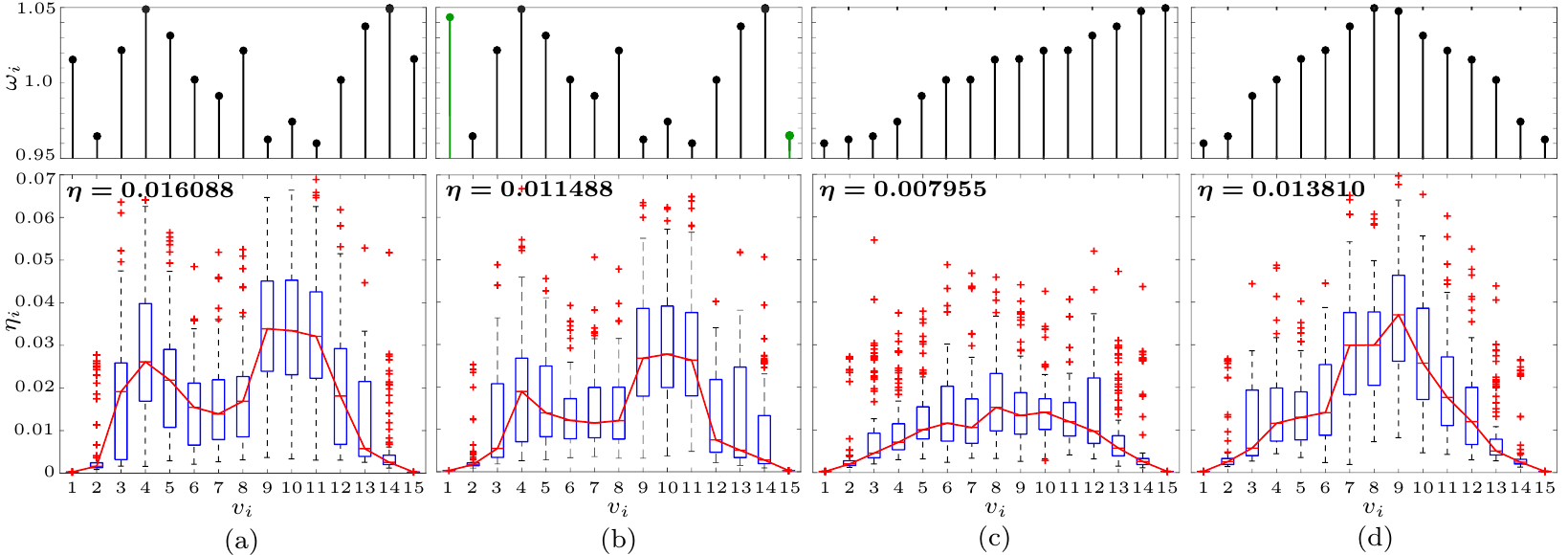}
	\caption{Boxplot of $\eta_i$ per node (bottom) and the corresponding natural frequency of the Kuramoto oscillator (top), considering the chain graph in Fig. \ref{fig:netchain}. (a) Random draw from $\omega_i\sim\mathcal{N}(1,0.03)$; (b) the same random draw of (a) with the natural frequencies $\omega_1$ and $\omega_{15}$ altered; (c) random draw in (a) sorted in ascending order; and (d) a unimodal arrangement of random draw in (a). The $\eta_i$ median is traced by a red line to guide the eyes. Simulations carried out for $\sigma_w=0.01$ and $1$, as well as $\sigma_v=0.01$ and $1$, yielded similar results.}
	\label{fig:kuramotormse}       
\end{figure*}

Guan \textit{et al.} show, through their proposed metric, that the observability level of a particular node seems to (intuitively) decrease with the increase of the path length between this node and a particular set of sensor nodes \cite{Guan2018}. Such analysis is directly related to the network topology and, considering the chain network setup in Fig. \ref{fig:netchain}, one might argue that the expected ``least observable node'' from  $\mathcal S=\{v_1,v_{15}\}$ would be the central node $v_8$. However, the numerical setup from which the authors of Ref. \cite{Guan2018} draw this particular conclusion assume that all nodal dynamics are equal (i.e. every node is governed by the same ordinary differential equations with the same parametric values).

But if the nodal dynamics were heterogeneous (i.e. a network with Kuramoto oscillators of different natural frequencies $\omega_i$ along the network structure), would $v_8$ still be the least observable one? It is known that the location of oscillators (with different $\omega_i$) in a network affects the synchronization process\cite{Boccaletti2002,Dorfler2014}, but how does it affect observability? For instance, Refs. \cite{Whalen2015,Haber2017} showed through ``small motifs'' that the optimal $\mathcal S$ depends not only on the ordinary differential equations that govern the system behaviour but also on its specific parameters.

It is here argued that the observability that results from the interplay between dynamics and topology can be understood assessing the performance of the PF estimates for different networks that share the same topology, but different dynamics. Figure~\ref{fig:kuramotormse} presents the attained filter estimates $\eta_i$ for different allocations of $\omega_i$ over the chain network in Fig. \ref{fig:netchain}.

Despite all networks in Figs. \ref{fig:kuramotormse}a,c,d being composed by the same Kuramoto oscillators, we see that their \textit{placing order} along the chain graph affects the PF performance in different manners. Depending on the oscillators location, the PF worst estimate shifts away from $v_8$, presenting different $\eta_i$ distributions over the network. For instance, in Figs. \ref{fig:kuramotormse}a,b we see a bimodal distribution of $\eta_i$, while in Fig. \ref{fig:kuramotormse}c,d there is a somewhat unimodal distribution. There is an implication between the degree of observability and the graph topology and dynamic interactions that seems to be, likewise in synchronization analysis \cite{Boccaletti2002}, related to the degree of mismatch between the dynamics of an oscillator and its neighbours. Adopting the convention in Ref. \cite{Boccaletti2002}, we refer as \textit{soft transition} if there is a relatively small frequency mismatch in a neighbourhood, or as \textit{hard transition} otherwise.

In the following analyses, we \textit{conjecture} that the oscillators nodes chosen as sensor nodes not only have to be topologically central to the network, but also must have a dynamical behaviour that is representative to (similar to) the dynamics of the other oscillators. In a sense, node oscillators chosen as sensor nodes in a dynamical network should be central both topologically and dynamically.

Consider Fig. \ref{fig:kuramotormse}a. Note that, although the central node $v_8$ is the farthest from a topological sense, it does not convey the worst observability, i.e. the higher estimate error $\eta_i$. Following the conjecture, despite its topological distance, $\omega_8$ is very close to the sensor nodes natural frequencies $(\omega_1, \omega_{15})$. On the other hand, estimates $(\eta_4, \eta_{10})$ and their respective neighbourhood show worse performance than $\eta_8$, despite being relatively closer than $v_8$ to the sensor nodes. The dynamical behaviour, however, is not quite central, with a high frequency mismatch between the oscillators $(v_4,v_{10})$ and their respective nearest sensor node $(v_{1},v_{15})$. Interestingly, $\eta_4<\eta_{10}$ can be justified by the fact that the degree of frequency mismatch between the neighbourhood of $v_4$ and its closer sensor node $s_1$ is smaller compared to $v_{10}$ and $s_2$. From the point-of-view of observability, in some way, the presence of dynamical centrality compensates for the lack of topological centrality, and vice-versa. Nevertheless, $v_2$ still provides good observability (low $\eta_2$) despite having a high frequency mismatch between $(v_1,v_2)$. This might be attributed to the close topological proximity of these nodes on the network, relieving the dynamic difference effects. In Fig. \ref{fig:kuramotormse}b, by only adjusting the oscillators frequencies $\omega_1$ and $\omega_{15}$ at the chain ends to respectively closer frequencies of $\omega_4$ and $\omega_{10}$ (and neighbourhood), the estimate error $\eta_i$ at this sections greatly reduced, with $\eta$ being approximately $30\%$ smaller than in Fig. \ref{fig:kuramotormse}a. 

Albeit the estimation error $\eta$ is certainly related to the degree of observability conveyed by the respective choice of $\mathcal S$, it is important to note that results are also impacted by the intrinsic dynamical relations. For instance, certain placing orders of oscillators along the network chain can give rise to formation of synchronous cluster manifolds, which effectively reduce the system dimension when complete synchronization holds, leading to a better convergence of the estimator. Moreover, the numerical results are also affected by the tunned levels of process and measurement noise as well as the discretization model of the PF.

Figures \ref{fig:kuramotormse}c,d support this analysis with different arrangements. In Fig. \ref{fig:kuramotormse}c, the frequency mismatch over the network chain is relatively low, pointing to soft transitions that favours global synchronization behaviour in dynamical networks \cite{Boccaletti2002}. Thus, the tendency to form a single synchronous manifold for the whole network, rather than isolated clustered manifolds in Fig. \ref{fig:kuramotormse}a, might lead to observations of better quality for the sensor nodes and, hence, a lower overall $\eta$ as seen in Fig. \ref{fig:kuramotormse}c. Moreover, being the natural frequencies arranged in a homogeneous manner, the central node might have the worst estimates due to its large topological distance from both sensor nodes as well as the high dynamic behaviour differences between $v_8$ and $s_1$ or $s_2$. The $\eta_i$ is mostly unimodal, although some slight peaks in $\eta_6$ and $\eta_{10}$ might be attributed, respectively, to relative soft transitions on pairs $(v_6,v_7)$ and $(v_{10}, v_{11})$. Likewise, in Fig. \ref{fig:kuramotormse}d, the worst estimates are central, due to high topological distance to sensor nodes and high dynamic behaviour differences.

\begin{figure}
	\centering
	\includegraphics[width=0.65\columnwidth]{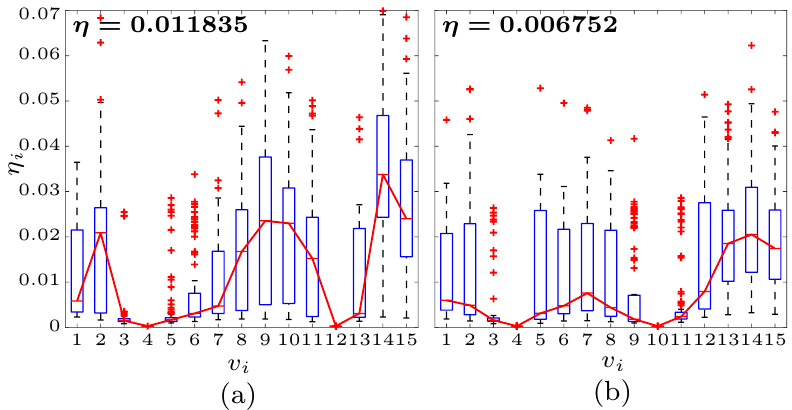}
	\caption{Boxplot of $\eta_i$ per node using the oscillators locations in Fig. \ref{fig:kuramotormse}a, but with different pairs of sensor nodes. (a)~$\mathcal{S}_2=\{v_4,v_{12}\}$; (b)~$\mathcal{S}_3=\{v_4,v_{10}\}$.}
	\label{fig:kuramotormse_newsensors}       
\end{figure}

Using the oscillators location along the chain graph of Fig. \ref{fig:netchain}, we show in Fig. \ref{fig:kuramotormse_newsensors} how the PF performance changes with the sensor locations. For instance, in Fig. \ref{fig:kuramotormse_newsensors}a, we set a pair of sensor nodes which provide the smallest path length from all other nodes in the network, while, in Fig. \ref{fig:kuramotormse_newsensors}b, we set $(v_4,v_{10})$ as the sensor nodes. Since $\eta$ is smaller in Fig. \ref{fig:kuramotormse_newsensors}b, we argue that choosing a set of sensor nodes that is not only central to the network topology, but also ``dynamically central'' (representative of its neighbourhood dynamics) seems to convey better degrees of observability. At least for the Kuramoto dynamical network, presence of synchronous manifolds, dynamical correlations and topological proximity seem to affect the network system observability. Nevertheless, the interplay between synchronization, observability and dynamics has already been addressed in Ref. \cite{Letellier2010}, with systems of lower dimensionality. The work pointed to a \textit{not exclusive} dependence between synchronization and observability, a remark that seems to be also relevant in high-dimensional systems.

\begin{figure}
	\centering
	\includegraphics[width=0.7\columnwidth,height=4cm]{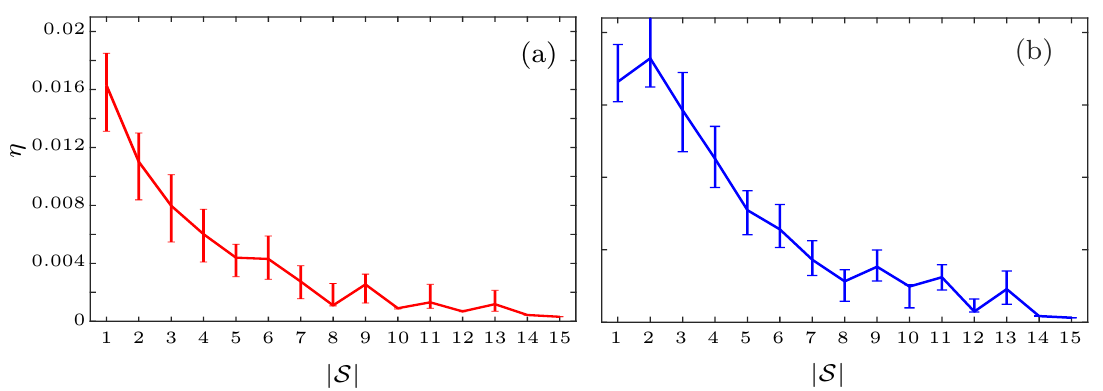}
	\caption{PF performance index $\eta$ per number of sensor nodes (allocated as detailed in text), for oscillators locations displayed in: (a)~Figs.~\ref{fig:kuramotormse}a (top), (b)~Figs.~ \ref{fig:kuramotormse}d (top). Error bars show the interquartile range.}
	\label{fig:numberofsensors}       
\end{figure}

Finally, in Fig. \ref{fig:numberofsensors}, we investigate the improvement in the PF performance $\eta$ by gradually increasing the number of sensor nodes. Sensor nodes are allocated in such a way that the sum of the path lengths between every single node and the closest sensor node is minimum. As the number of sensors increase, the PF usually benefits from more valuable information and, consequently, $\eta$ has a tendency to decrease in an exponential manner. This is related to Haber \textit{et al.} \cite{Haber2017} conclusion that there is an intrinsic obstacle when trying to infer the network states from a small number of sensors. Interestingly, in some instances, $\eta$ increases after adding a sensor node. Due to the sensor nodes allocation method, from an even number of nodes to an odd number, not only the number of sensors increase but their positions change as well. Thus, increasing the number of sensor nodes does not guarantee better average observability. This result is also present in Fig. \ref{fig:obsmetrics}, where some sensor sets of cardinality $2$ provide better observability than a set of cardinality $3$. For instance, Fig. \ref{fig:numberofsensors}b shows clearly that the PF performance, and the network observability, are better when not including the central node as a sensor node, which always happens when there is an odd number of sensors. Interestingly, the higher the observability, the lower the median and the interquartile range of $\eta$\textemdash indicating that the PF estimates are more reliable. 

\subsection{Network of R\"ossler systems}
\label{sec:rosslerresults}

\begin{figure*}
	\centering
	\includegraphics[width=\textwidth]{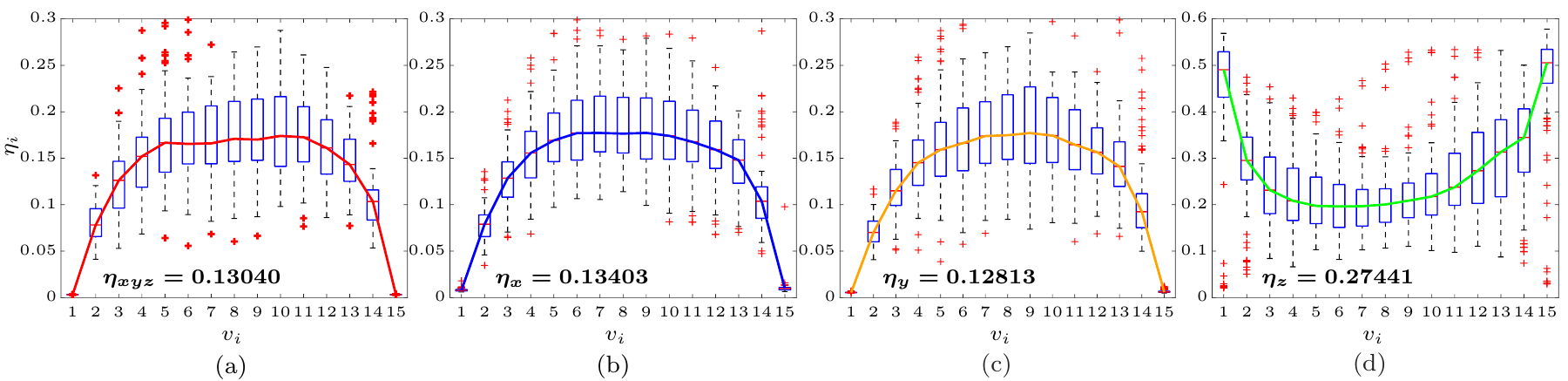}
	\caption{Boxplot of $\eta_i$ per node, for a chain graph of R\"ossler systems (coupled by the $y$ variable) with different sets of sensor nodes: (a) $\mathcal{S}_1=\{x_1,y_1,z_1, x_{15},y_{15},z_{15})\}$; (b) $\mathcal{S}_2=\{x_1,x_{15}\}$; (c) $\mathcal{S}_3=\{y_1,y_{15}\}$; (d)~$\mathcal{S}_4=\{z_1,z_{15}\}$. Simulations carried out for $\sigma_w=0.001$ and $0.0001$, as well as $\sigma_v=0.01$ and $0.001$, yielded similar results. For $\sigma_w\geq1$, the process noise degenerates the filter dynamics and it diverges.}
	\label{fig:rosslerrmse}       
\end{figure*}

Consider a R\"ossler dynamical network of $N=15$ nodes ($n=3N$ states) where measures are taken independently on the chain extremities ($\mathcal{S}=\{v_1,v_{15}\}$, see Fig. \ref{fig:netchain}). In this network system, each node is composed by three state variables whose observability rank is $y\rhd x\rhd z$, with $y$ and $x$ being  good variables to reconstruct the dynamics, and $z$ a poor one \cite{Letellier2005}. In this section, based on the PF performance, we investigate how the observability of nodal dynamics affects the network observability. Thus, we compare the numerical results for the following sets of sensor nodes: (i)~all the variables of a sensor node are measured, i.e. $\mathcal{S}_1=\{(x_1,y_1,z_1), (x_{15},y_{15},z_{15})\}$; and (ii)~only one variable of a sensor node is measured i.e. $\mathcal{S}_2=\{x_1,x_{15}\}$, $\mathcal{S}_3=\{y_1,y_{15}\}$ and $\mathcal{S}_4=\{z_1,z_{15}\}$.

Fig. \ref{fig:rosslerrmse} presents the attained filter estimates $\eta_i$ for each of the described sets of sensor nodes. As described in Section \ref{sec:pfrossler}, the implemented R\"ossler system network has a soft transition in parameter $a_i$ along the network chain. Although it is not our goal to evaluate the effects of topology in network observability in this section, we note that the central node $v_8$ has the highest $\eta_i$ in every occasion, just as seen in Fig. \ref{fig:kuramotormse}c.

Clearly, the observability order of R\"ossler system is carried over to its high-dimensional network, as $\eta_y < \eta_x < \eta_z$. The poor performance of $\eta_{z}$ is likely associated with the little information (poor observability) that the $z$ variable has regarding the system cyclic behaviour, which hampers state reconstruction, especially when $z_i \approx 0$. Curiously, as the path length from a node estimate to a sensor node increases, $\eta_{z_i}$ reduces rather than increasing as in the other examples. This counter-intuitive phenomenon is yet to be explained. Despite this, all estimates $\eta_{z_i}$ in Fig.~\ref{fig:rosslerrmse}d are always inferior compared to the others. 
Moreover, the result that $\eta_y < \eta_{xyz}$, although only slightly, brings us to a recurring conclusion found in Ref. \cite{Portes2016a} that, in some cases, to measure all variables is not necessarily better than to measure the best one in terms of observability.

The attained effects of a multidimensional node over the network observability in this study can be studied under a graph point-of-view \cite{Aguirre2018}. A single multidimensional node can be decomposed in multiple additional and interconnected nodes (states) as illustrated in Fig.~\ref{fig:rosslergraph}. In the case of a R\"ossler system, based on system model \eqref{eq:rosslernetwork}, a single node $v_i$ can be ``unfolded'' into three nodes: $(v_{x_i}, v_{y_i}, v_{z_i})$, where their interconnections can be deduced from the system equation following Lin's structural controllability approach \cite{Lin1974a}. The reader is referred to Refs. \cite{Letellier2018,Aguirre2018} for further details. Nevertheless, notice that $v_{z_i}$ has a direct path to $v_{x_i}$ that is weighted by the state $z_i$. Thus, when $z_i\rightarrow 0$, this path vanishes and $ v_{z_i}$ loses information of  $v_{x_i}$ and, consequently, of all network. This behaviour is probably another way of understanding the loss of performance in Fig. \ref{fig:rosslerrmse}d.

\section{Conclusion}
\label{sec:conc}


In this paper, we develop a particle filtering framework as a way to assess observability properties of a dynamical network, where each node is composed by an individual dynamical system. This is motivated by the fact that the quality of the filter estimation depends on the ``level'' of observability provided by the signal.
Firstly, we show the PF effectiveness to quantify the observability level in a network by comparing this framework to a traditional observability metric in a system of lower dimensional order. Secondly, we show, in a network of $15$ and $45$ states (composed by Kuramoto oscillators and R\"ossler systems, respectively), that a good choice of sensors is related not only to the topological position of a sensor node in a graph, but also to its dynamic behaviour\textemdash that is, the precise values of the model parameters. These results corroborate with Ref. \cite{Haber2017}. We conjecture that the sensor node dynamics must share some dynamical affinity to its neighbours in order to be representative of their behaviour and, thus, carry relevant information to the measurement signals. Thirdly, we investigate how the choice of an internal measured variable of a sensor node can affect the overall network observability. As shown in a R\"ossler network, the PF performance, and thus the observability, is better when measures are taken on variables of higher observability of the uncoupled node dynamics.



This work might provide insight for future researches on dynamical networks and the interplay between topology and dynamics. The detailed framework of Bayesian filtering as a performance measure of observability can be used as a validation platform over different benchmarks of networks in order to compare novel observability metrics in literature. This framework, however, is dependent on an appropriate tuning of the filter internal parameters, such as the process and measurement noise levels. This feature is also present in other observability metrics that relies on the fitting error of a state reconstructor \cite{Guan2018,Carroll2018}.



\appendix

\section{Observability and particle filtering \\ of chaotic systems benchmarks}
\label{sec:chaoticsystems}

This section aims to validate, along with Section \ref{sec:pfbenchmark} results, the PF framework effectiveness\textemdash when well tuned\textemdash as an observability metric. To this purpose, we investigate the observability degree $\delta_o$ and the overall PF estimate error $\eta$ on five 3-dimensional chaotic systems benchmarks explored in \cite{Aguirre2011a}. Table \ref{tab:chaoticbench} presents the observability rank derived by $\delta_o$ and $\eta$ (for a 100 Monte Carlo iterations) when variables $x$, $y$ and $z$ are recorded one at a time. For details on the chaotic systems ordinary differential equations as well as chosen parameters, we refer the reader to Ref. \cite{Aguirre2011a}. 

All numerical integrations were performed using a fourth-order Runge Kutta algorithm for a simulation time of $200$ s, with time step $0.01$ s (except for the cord system, which we set $0.001$ s). For all chaotic benchmarks, the PF framework follows Algorithm \ref{alg:pfkuramoto}. Since in some chaotic systems the state variables magnitude range are significantly different, random noise $\bm w_k\in\R^3$ and $\bm v_k\in\R^3$ are independent, but not identical, random variables following a normal distribution $\mathcal N(0,\Sigma)$. The diagonal entries of $\Sigma_w$ and $\Sigma_v$ are shown in Table \ref{tab:chaoticbench}. In order to tune the PF for each chaotic system benchmark, $\Sigma_w$ and $\Sigma_v$ were adjusted following a model validation testing described in Ref. \cite[Section 7.6.4]{Candy2016a}, which is grounded on a whiteness test analysis for a PF framework. This model validation test was not applied in Section \ref{sec:setup} since it is not quite robust for high-dimensional systems. 

\begin{table}
	\caption{Comparison of the observability degree and the PF estimation error of chaotic systems benchmarks.}
	\label{tab:chaoticbench}       
	\centering
	\begin{tabular}{ll|ll|ll}
		\toprule[2pt]
		Chaotic system & & $\delta_o$\footnotemark[1] & $\eta$ [Q1 Q2 Q3]\footnotemark[2] & $\Sigma_w$ & $\Sigma_v$ \\
		\toprule[2pt]
		 & x & 0.022 & [0.68 1.02 1.45] & 0.01 & 0.5 \\
		R\"ossler system & y & 0.133 & [0.62 0.96 1.39] & 0.01 & 0.5 \\
		 & z & 0.0063 & [1.29 1.91 2.92] & 0.01 & 0.5 \\
		 &   & $y\rhd x\rhd z$ & $y\rhd x\rhd z$ & & \\
		\midrule[0.5pt]
		& x & 0.0104 & [0.12 0.20 0.40] & 0.001 & 0.05 \\
		Cord system & y & 0.0005 & [0.27 0.39 0.59] & 0.01 & 0.5 \\
		& z & 0.0005 & [0.28 0.42 0.62] & 0.01 & 0.5 \\
		&   & $x\rhd y\approx z$ & $x\rhd y\approx z$ & & \\
		\midrule[0.5pt]
		& x & -- & [0.24 0.37 0.54] & 0.05 & 0.6 \\
		Double-scroll\footnotemark[3] & y & -- & [0.43 0.64 0.99] & 0.01 & 0.2 \\
		& z & -- & [0.22 0.35 0.56] & 0.05 & 0.6 \\
		&   & $x\rhd z\rhd y$ & $z\approx x\rhd y$ & & \\
		\midrule[0.5pt]
		& x & $6.5\times 10^{-6}$ & [2.56 3.64 5.14] & 1 & 2 \\
		Lorenz system & y & $8.2\times 10^{-6}$ & [2.13 2.85 3.88] & 1 & 2 \\
		& z & $2.2\times 10^{-5}$ & [8.73 17.0 33.6] & 1 & 2 \\
		&   & $z\rhd x\rhd y$ & $y\rhd x\rhd z$ & & \\
		\midrule[0.5pt]
		& x & $0.0142$ & [4.07 6.29 8.25] & 0.005 & 0.25 \\
		Lorenz'84 & y & $0.0009$ & [1.99 2.68 3.83] & 0.01 & 0.5 \\
		& z & $0.0014$ & [2.05 2.80 4.07] & 0.01 & 0.5 \\
		&   & $x\rhd z\approx y$ & $y\approx z\rhd x$ & & \\
		\toprule[2pt]
	\end{tabular}

	\raggedright
	\footnotesize{\footnotemark[1]Observability indices were computed in Refs. \cite{Aguirre2008,Aguirre2011a}.} \\
	\footnotesize{\footnotemark[2]First, second (median) and third quartiles ($\times 10^{-2}$).} \\
	\footnotesize{\footnotemark[3]Observability indices computed for the double-scroll attractor yielded measures that were practically zero (smaller than $10^{-15}$) \cite{Aguirre2008}. We referred to the computed indices in Ref. \cite{Aguirre2011a} to sort the variables from best to worst observability.}
\end{table}

All results are statistically consistent, as shown by the quartile ranges. In the chaotic systems of uttermost observability properties, the R\"ossler and Cord systems, both metrics $\delta_o$ and $\eta$ ranked the variables from best to worst degree of observability in the same order. In the double-scroll attractor, both metrics indicate $y$ as the variable that conveys worse observability\textemdash which is expected since this variable does not distinguish the three equilibrium points. Despite the disagreement regarding the $x$ and $z$ variables, both convey good observability. The Lorenz system provides an interesting result. It is known that the $z$ variable does not distinguish between the two different equilibrium points on each of the attractor wings. Since this information is not detectable by a local measure, such as $\delta_o$, it ranks $z$ as variable of good observability\cite{Letellier2002}. The PF estimation error, on the other hand, detects this global property, inferring that $z$ is the variable of worst observability. This result has also been detected using the training error of a reservoir computer\cite{Carroll2018}. Finally, albeit the degree of observability conveyed by $x$ in the Lorenz'84 is different for each method, both recognizes that the $y$ and $z$ variables provides similar results.


\section*{Acknowledgements}
This study was financed in part by the Coordena\c{c}\~ao de Aperfei\c{c}oamento de Pessoal de N\'ivel Superior - Brasil (CAPES) - Finance Code 001, and by the Conselho Nacional de Desenvolvimento Cient\'ifico e Tecnol\'ogico - Brasil (CNPq).

\bibliographystyle{apalike}
\bibliography{library}

\end{document}